# Harnessing high-dimensional hyperentanglement through a biphoton frequency comb

Zhenda Xie[1,2], Tian Zhong[3], Sajan Shrestha[2], XinAn Xu[2], Junlin Liang[2], Yan-Xiao Gong[4], Joshua C. Bienfang[5], Alessandro Restelli[5], Jeffrey H. Shapiro[3], Franco N. C. Wong[3], and Chee Wei Wong[1,2]

[1] *Mesoscopic Optics and Quantum Electronics Laboratory, University of California, Los Angeles, CA 90095*

[2] *Optical Nanostructures Laboratory, Columbia University, New York, NY 10027*

[3] *Research Laboratory of Electronics, Massachusetts Institute of Technology, Cambridge, MA 02139*

[4] *Department of Physics, Southeast University, Nanjing, 211189, People's Republic of China*

[5] *Joint Quantum Institute, University of Maryland and National Institute of Standards and Technology, Gaithersburg, Maryland 20899, USA*

**Quantum entanglement is a fundamental resource for secure information processing and communications, where hyperentanglement or high-dimensional entanglement has been separately proposed for its high data capacity and error resilience. The continuous-variable nature of the energy-time entanglement makes it an ideal candidate for efficient high-dimensional coding with minimal limitations. Here we demonstrate the first simultaneous high-dimensional hyperentanglement using a biphoton frequency comb to harness the full potential in both energy and time domain. Long-postulated Hong-Ou-Mandel quantum revival is exhibited, with up to 19 time-bins and 96.5% visibilities. We further witness the high-dimensional energy-time entanglement through Franson revivals, observed periodically at integer time-bins, with 97.8% visibility. This qudit state is observed to simultaneously violate the generalized Bell inequality by up to**

**10.95 standard deviations while observing recurrent Clauser-Horne-Shimony-Holt *S*-parameters up to 2.76. Our biphoton frequency comb provides a platform for photon-efficient quantum communications towards the ultimate channel capacity through energy-time-polarization high-dimensional encoding.**

Increasing the dimensionality of quantum entanglement is a key enabler for high-capacity quantum communications and key distribution [1, 2], quantum computation [3] and information processing [4, 5], imaging [6], and enhanced quantum phase measurement [7, 8]. A large Hilbert space can be achieved through entanglement in more than one degree of freedom (known as hyperentanglement [2, 7, 9]), where each degree of freedom can also be expanded to more than two dimensions (known as high-dimensional entanglement). The high-dimensional entanglement can be prepared in several physical attributes, for example, in orbital angular momentum [1, 10-12] and other spatial modes [13-15]. The drawback of these high-dimensional spatial states is complicated beam-shaping for entanglement generation and detection, which reduces the brightness of the sources as the dimension scales up, and complicates their use in optical-fiber-based communications systems. In contrast, the continuous-variable energy-time entanglement [16-22] is intrinsically suitable for high-dimensional coding and, if successful, can potentially be generated and be communicated in the telecommunication network. However, most studies focus on time-bin entanglement, which is discrete-variable entanglement with typical dimensionality of two [23-25]. Difficulties in pump-pulse shaping and phase control limit the dimensionality of the time-bin entanglement [26], and high-dimensional time-bin entanglement has not been fully characterized because of the overwhelmingly complicated analyzing interferometers. On the other hand, a biphoton state with a comb-like spectrum could potentially serve for high-dimensional entanglement generation and take full advantage of the continuous-variable energy-time subspace. Based on this state, promising applications have been proposed for quantum computing, secure wavelength-division multiplexing, and dense quantum key distribution [3, 27, 28]. A phase-coherent biphoton frequency comb (BFC) is also known for its mode-locked behavior in its second-order correlation. Unlike classical frequency combs, where mode-locking directly relies on phase coherence over individual comb lines, the mode-locked behavior of a BFC is the representation of the phase coherence of a biphoton wavepacket over comb-line pairs, and results in periodic recurrent correlation at different time-bins [29, 30]. This time correlation feature can be characterized through quantum interference when passing the BFC through an unbalanced Hong-Ou-Mandel (HOM)-type interferometer [31]. A surprising revival of the correlation dips can be observed at time-bins with half the period of the BFC revival time. However, because of the limited type-I collinear spontaneous parametric downconversion (SPDC) configuration in the prior studies [29], post-selection was necessary for the BFC generation where the signal and idler photons are indistinguishable, limiting the maximum two-photon interference to 50 %.

Here we achieve high-dimensional hyperentanglement through a biphoton frequency comb. The high-dimensional hyperentanglement of the BFC is fully characterized with four observations. First, the state is prepared at telecommunications wavelengths, without the necessity of post-selection, by using a type-II high-efficiency periodically-poled $KTiOPO_4$ (ppKTP) waveguide together with a fiber Fabry-Perot cavity (FFPC). Because of the type-II phase matching, signal and idler photons can be separated efficiently by a polarizing beamsplitter (PBS), allowing deterministic BFC generation, as first proposed theoretically by one of the authors [29], to be observed experimentally for the first time. Revival dips with $\approx$ 96.5 % visibility from two-photon interference in a HOM-type interferometer are observed recurrently for the first time, which reveals correlation features in the time bins of the BFC. Second, second-order frequency correlation and anticorrelation, scanned across the full span of the frequency bins by narrowband filter pairs, shows the good fidelity of the frequency-bin entangled state. Finally, we confirm the generation of high-dimensional energy-time entanglement using a Franson interferometer, which can be regarded as a generalized Bell inequality test. For the first time, Franson interference fringes are observed to revive periodically at different time-bin intervals with visibilities up to 97.8 %. Based on these three measurements, we encode extra qubits in the BFC photon pairs by mixing them on a 50:50 fiber beam splitter (FBS). We witness the hyperentanglement by simultaneous polarization and Franson interferometer analysis, demonstrating a generalized Bell inequality in two polarization and four time-bin subspaces up to 10.95 and 8.34 standard deviations respectively. The Clauser-Horne-Shimony-Holt (CHSH) *S*-parameters are determined for the polarization basis across the different time-bins with a maximum up to 2.76.

Figure 1(a) shows our experimental scheme. The SPDC entangled photon pairs are generated by a high-efficiency type-II ppKTP waveguide, described in detail in [32]. The frequency-degenerate SPDC phase matching is designed for 1316 nm output with a bandwidth of about 245 GHz. The type-II BFC is generated by passing the SPDC photons through a fiber Fabry-Perot cavity (Micron Optics[i]) with the signal and idler photons in orthogonal polarizations (*H*: horizontal and *V*: vertical). A BFC state is expressed as:

$$|\psi\rangle = \sum_{m=-N}^{N} \int d\Omega f(\Omega - m\Delta\Omega) \hat{a}_H^\dagger(\omega_p/2 + \Omega) \hat{a}_V^\dagger(\omega_p/2 - \Omega)|0\rangle, \qquad (1)$$

where $\Delta\Omega$ is the spacing between the frequency bins, i.e., the free spectral range of the FFPC in rad/s; $\Omega$ is the detuning of the SPDC biphotons from their central frequency; and the state's spectral amplitude, $f(\Omega - m\Delta\Omega)$, is the single frequency bin profile, defined by the Lorentzian-shape transmission of the FFPC with full-width half-maximum (FWHM) of $2\Delta\omega$.

$$f(\Omega) = 1/[(\Delta\omega)^2 + \Omega^2], \qquad (2)$$

The signal and idler photons are separated with 100 % probability by a polarizing beamsplitter or, in other words, the BFC is thus prepared without post-selection. The temporal wave function of the BFC can be written as:

$$|\psi_t\rangle = \int d\tau \exp(-\Delta\omega|\tau|) \frac{\sin[(2N+1)\Delta\Omega\tau/2]}{\sin(\Delta\Omega\tau/2)} \hat{a}_H^\dagger(t) \hat{a}_V^\dagger(t+\tau)|0\rangle, \qquad (3)$$

---

[i] The identification of any commercial product or trade name does not imply endorsement or recommendation by the National Institute of Standards and Technology.

where the exponential decay is slowly varying because of small $\Delta\omega$, and thus the temporal behavior of the BFC is mainly determined by the term $\frac{\sin[(2N+1)\Delta\Omega\tau/2]}{\sin(\Delta\Omega\tau/2)}$, with a repetition time $T = 2\pi/\Delta\Omega$.

The FFPC has free spectral range (FSR) and bandwidth of 15.15 GHz, 1.36 GHz, respectively. The repetition period $T$ of the BFC is about 66.2 ps. The FFPC is mounted onto a thermoelectric cooling sub-assembly with minimized stress to eliminate polarization birefringence and with $\approx$ 1 mK high-performance temperature control. From our measurements, there is no observable polarization birefringence in the FFPC, and thus the signal and idler photons have the same spectrum after they pass through the cavity. Due to the type-II configuration, there is no probability that both photons propagate in the same arm of the HOM interferometer − this configuration yields a potential maximum of 100 % visibility of the two-photon interference.

The pump is a Fabry-Perot laser diode that is stabilized with self-injection-locking, through double-pass first-order-diffraction feedback using an external grating (for details, see Supplementary Information and Methods). It is passively stabilized with modest low-noise current and temperature control for single-longitudinal-mode lasing at 658 nm, with a stability of less than 2 MHz within 200 s (measured with the Franson interference experiment, see Supplementary Information). We first match the FFPC wavelength with the pump via temperature tuning and second harmonic generation (SHG) from a frequency-stabilized distributed feedback (DFB) reference laser at 1316 nm. The SHG is monitored with a wavemeter (WS-7, HighFinesse) to 60 MHz accuracy, allowing the tuning of the laser diode current and first-order-feedback diffraction angle to match the FFPC for the BFC generation. The BFC spectrum is further cleaned by a fiber Bragg grating (FBG) with a circulator, and a free-space long-pass filter that blocks the residual pump light. The FBG is chosen with a bandwidth of 346 GHz, larger than the 245 GHz phase-matching bandwidth, and is simultaneously temperature-controlled to match the central wavelength of the SPDC. A polarizing beamsplitter separates the orthogonally polarized signal and idler photons.

We first characterize the mode-locked behavior using a HOM interferometer. The signal and idler photons are sent through different arms of the interferometer. A fiber polarization controller in the interferometer's lower arm controls the idler photon polarization, so that the two photons have the same polarization at the FBS. An optical delay line consisting of a prism and a motorized long-travel linear stage (DDS220, Thorlabs) is used to change the relative timing delay between the two arms of the HOM interferometer. The position-dependent insertion loss of the optical delay line is measured to be less than 0.02 dB throughout its entire travel range of 220 mm. The coincidence measurements are performed with two InGaAs single-photon detectors $D_1$ (with ≈ 20 % detection efficiency and 1 ns gate width) and $D_2$ (with ≈ 25 % detection efficiency and 3 ns gate width). $D_1$ is triggered at 15 MHz, and its output signal is used to trigger $D_2$. As a result, coincidences can be detected directly from the $D_2$ counting rate if the proper optical delay is applied to compensate for the electronic delay. Taking into account the waveguide-to-fiber coupling and transmission efficiencies of optical components, the overall signal and idler detection efficiencies are estimated to be $\eta_s = 0.92$ % and $\eta_i = 1.14$ % respectively. Figure 1(b) shows the experimental results by scanning the optical delay between the two arms of the HOM interferometer from -320 ps to 320 ps, with a pump power of 2 mW. At this power level the generation rate is about $3.3 \times 10^{-3}$ pairs per gate, where multi-pair events can be neglected. We obtain revival dips for the coincidence counting rate $R_{12}$ at the two outputs of the HOM interferometer (single-photon rates are shown in the Supplementary Information). The spacing between the dips is 33.4 ps, which matches half the repetition period of the BFC, and agrees with our theory and numerical modeling (see Supplementary Information I for details). We also note that our prior analysis [30] involved a movable beam splitter for both beams (for a *T* recurrence) while in our measurement setup only the idler beam is delayed (for a *T/2* recurrence), supported by the same type of analysis and physical interpretation. The visibility of the dips decreases exponentially [see right insert in Figure 1(b)] due to the Lorentzian lineshape spectra of the SPDC individual photons after they pass through the FFPC. A zoom-in of the dip around the zero delay point is shown in the insert of Figure 1(b). The maximum visibility is measured to be 87.2 ± 1.5 %, or 96.5 % after subtracting the accidental coincidence counts ( ≈ 17.3 per 70 sec). Its base-to-base width is fitted to be 3.86 ± 0.30 ps, corresponding to a two-photon bandwidth of 259 ± 20 GHz, which agrees with the expected 245 GHz phase-matching bandwidth. More details are provided in Supplementary Information V. Considering the measured bin spacing $\Delta\Omega/2\pi$ of 14.98 GHz, 17 frequency bins are generated in our measurements within the phase matching bandwidth, equivalent to ≈ 4 quantum bits per photon for high-dimensional frequency entanglement.

To test the purity of the high-dimensional frequency bin entanglement, we further measure the correlation between different frequency bins for the signal and idler photons. Each pair of the frequency-bins is filtered out by a set of narrow band filters for the signal and idler photons. As shown in Figure 2, each filter is composed of a FBG centered at 1316 nm with 100 pm FWHM and an optical circulator. Both FBGs are embedded in a custom-built temperature-controlled housing for fine spectral tuning. The coincidence counting rate is recorded while the filters are set at different combinations of the signal-idler frequency bin basis. The tuning range is bounded by the maximum tuned FBG temperature (up to about 100°C currently) and in this case 9 frequency bins can be examined for the signal and idler biphoton. We denote these bins with numbers #-4 to #4, with #0 standing for the central bin, as shown in Figure 2(b). We see that high coincidence counting rates are measured only when the filters are set at the corresponding positive-negative frequency bins, according to Equation 1. The suppression ratio between the corresponding and non-corresponding counting rate exceeded 13.8 dB for adjacent bins (including ≈ 2 % leakage from the band pass filters), or 16.6 dB for the non-adjacent bins. We note that, through conjugate state projection [20], observation of both the HOM recurrence and frequency correlation supports the entangled nature of the BFC.

To further characterize the energy-time entanglement of the BFC, we build an interferometer to examine Franson interference as schematically shown in Figure 3(a). The Franson interferometer [16, 33] comprises two unbalanced Mach-Zehnder interferometers (detailed in Methods and Supplementary Information III), with imbalances $\Delta T_1$ and $\Delta T_2$. Usually, Franson interference comes from two indistinguishable two-photon events: both signal and idler photons take the long arm (L-L) and both photons take the short arm (S-S). This interference only happens when $\Delta T \equiv \Delta T_1 - \Delta T_2 = 0$ to within the single photon coherence length, which is 1.81 ps in our case. Because of the high-dimensional frequency entanglement, however, Franson interference can be observed at different time revivals. As discussed before, the BFC features a mode-locked revival in temporal correlation, with repetition period of $T$. Therefore such Franson interference revival can be observed for integer temporal delays $\Delta T = NT$ where $N$ is an integer including 0. Detailed theory and modeling are illustrated in Supplementary Information II for the revival of the Franson interference. The experimental setup is shown in Figure 3(b). The signal and idler photons are sent to two fiber-based interferometers (*arm1* and *arm2*) with imbalances $\Delta T_1$ and $\Delta T_2$, respectively. Details on the pump laser and interferometer setup stabilization are described in Supplementary Information III and IV, and in the Methods. Each arm is formed by double-pass temperature-stabilized Michelson interferometers, and two of the output ports of the fiber 50:50 beamsplitter are spliced onto two Faraday mirrors, with single photons being collected from the reflection. Thus they work effectively as a Mach-Zehnder interferometer, and the polarization instability inside the fibers is accurately self-compensated. Following the general requirement for Franson interference, the $\Delta T_1$ and $\Delta T_2$ time differences (5 ns in our setup) are tuned to be much larger than the single-photon coherence time and the timing jitter of the single-photon detectors. The relative timing between the single photon detection gates are programmed such that events are only recorded for the L-L and S-S events. Both arms are mounted on aluminum housings with $\approx 1$ mK temperature control accuracy. Enclosures are further used to seal the interferometers with additional temperature control for isolating mechanical and thermal shocks. To detect the Franson interference at different temporal revivals, we include a free-space delay line on *arm2*, which gives us the capability to study the Franson interference with possible detuning $\Delta T$ up to 360 ps, bounded by the free-space delay line travel and for up to six positive-delay time bins.

Here we are interested in the regime of $\Delta T \geq 0$, because of the symmetry of the signal and idler. In particular, this allows us to examine the interference visibility over a larger time frame, for the same stage travel range. With $\Delta T_2$ fixed at each point, the phase sensitive interference is achieved by fine-control of the *arm1* temperature, which tunes $\Delta T_1$. As shown in Figure 4, the revival of Franson interference is only observed exactly at the periodic time bins $\Delta T = NT$, while no interference is observed for other $\Delta T$ values between these bins. The period of the revival time bins is 66.7 ps, which corresponds exactly to the round-trip time of the FFPC ($2\pi/\Delta\Omega$). The visibility of the interference fringes are measured to be 94.2 %, 89.3 %, 79.7 %, 71.1 %, 51.0 % and 43.6 %, or, after subtracting the accidental coincidence counts, 97.8 %, 93.3 %, 83.0 %, 74.1 %, 59.0 %, and 45.4 % respectively. The visibility decreases because of the non-zero linewidth of each frequency bin and is captured in our theory (see Supplementary Information II). We understand this as a generalized Bell inequality violation for a high-dimensional state at 4 time bins from the center. We only measured for $\Delta T \geq 0$, but could expect the violation of the generalized Bell inequality at the other three inverse symmetric time bins of $\Delta T$ at -66.7 ps, -133.4 ps, and -200.1 ps based on symmetry with the three positive time bins observed.

Figure 5 next shows the high-dimensional hyperentanglement by mixing the signal and idler photons on a 50:50 FBS (detailed in Supplementary Information VI). In the experiment, hyperentanglement is generated based on the HOM-interference setup fixed at the central dip, but with polarization adjusted so that the signal and idler photons are orthogonally polarized at the FBS for hyperentanglement generation. Two polarizers P1 and P2 are used for the polarization analysis, which is cascaded with the Franson interferometer for the energy time entanglement measurement at the same time. One set of half and quarter waveplates are placed before each polarizer to compensate the polarization change in the fiber after the FBS. An additional thick multi-order full waveplate is used in the lower arm and it can be twisted along its fast axis (fixed in the horizontal plane) to change the phase delay between the horizontally and vertically polarized light. Therefore, we successfully generated the following high-dimensional hyperentangled state when a coincidence is measured.

$$|\psi'\rangle = \sum_{m=-N}^{N} \int d\Omega f(\Omega - m\Delta\Omega)(|H, \omega_p/2+\Omega\rangle_1 |V, \omega_p/2-\Omega\rangle_2 + |V, \omega_p/2+\Omega\rangle_1 |H, \omega_p/2-\Omega\rangle_2) \quad (4)$$

Such hyperentanglement is tested using coincidence measurement by scanning P2 and the Franson interferometer while P1 is set at 45° or 90° and $\Delta T$ fixed at time bins #0 to #5. A 1.3 GHz self-differencing InGaAs single-photon detector [34, 35] is used as D1 to maximize the gated detection rate. As shown in Figure 5(b) and 5(c), we obtained clear interference fringes over both polarization and energy-time basis, with visibilities up to 96.7 % and 95.9 % respectively, after dark count subtraction of a 0.44 s$^{-1}$ rate. Analyzed within the two polarization and four time-bin subspaces, this corresponds to Bell violation up to 10.95 and 8.34 standard deviations, respectively. The CHSH $S$ parameter is observed for the polarization basis up to $S = 2.76$, with details on the photon statistics shown in Figure 5(d) and listed in Table 1 of the Supplementary Information VII.

In summary, we have demonstrated a high-dimensional hyperentanglement of polarization and energy-time subspaces using a BFC. Based on continuous-variable energy-time entanglement, unlimited qubits can be coded on the BFC by increasing the number of comb line pairs with plane wave pump. Here we show an example for bright BFC generation with 5 qubits per photon, which is from a type-II SPDC process in a ppKTP waveguide without post-selection. 19 Hong-Ou-Mandel dip recurrences with a maximum of 96.5 % visibility in a stabilized interferometer are observed. High-dimensional energy-time entanglement is proven by Franson interference, which can be considered as a generalized Bell inequality test. Revival of the Franson interference has been witnessed at the periodic time bins, where the time bin separation is the cavity round trip time, i.e., the revival time of the BFC. The interference visibility is measured up to 97.8 %. The generalized Bell inequality has been violated at 4 out of 6 measured time bins, or 7 time bins in total considering the symmetry. High-dimension hyperentanglement is further generated and characterized with high fidelity in both polarization and energy-time subspaces, with Bell inequality violations up to 10.95 and 8.34 standard deviations respectively, with the measured polarization CHSH *S*-parameters up to 2.76. It should be noted that the generation rate of the entangled biphoton frequency comb from our high-performance ppKTP waveguide exceeds 72.2 pairs/s/MHz/mW, by taking account of all detection losses and the 1.5 % duty cycle for the detection time, which is higher than that of cavity-enhanced SPDC using bulk crystals. If the propagation loss of the ppKTP waveguide can be further reduced, the waveguide can be put inside an optical cavity and the brightness of the BFC further enhanced. Such a bright high-dimensional hyperentangled BFC can encode multiple qubits onto a single photon pair without losing high photon flux, and thus further increase the photon efficiency with applications in dense quantum information processing and secure quantum key distribution channels.

**Methods**

**Stabilized self-injection-locked 658 nm pump laser:** For the Franson measurements, good long- and short-term stability of the 658 nm pump laser is required since this stability defines the coherence time of the two-photon state. This coherence time should be much longer than the path length difference ($\approx$ 5 ns in our Franson measurements). Hence we have designed and built a stabilized laser through self-injection-locking, in a configuration similar to the Littman-Metcalf cavity. With its single spatial mode, the longitudinal modes are selected through a diffraction grating (first order back into the diode; zeroth order as output) with an achieved mode-rejection ratio of more than 30 dB. Three temperature controllers are used to stabilize the doubly-enclosed laser system, and a low-noise controller from Vescent Photonics (D2-105) drives the laser diode. With our design, a 2 MHz stability is achieved for 200 second measurement and integration timescales, confirmed with the Franson interferometer. For longer timescales (12 hours), a wavelength meter indicates $\approx$ 100 MHz drift. More details on the characterization, setup, and design are illustrated in the Supplementary Information.

**Franson interferometer:** For our long-term phase-sensitive interference measurements, the fiber-based Franson interference needs to be carefully stabilized. Both interferometer arms are double-enclosed and sealed, and temperature-controlled with Peltier modules. Closed-loop piezoelectric control fine-tunes the *arm2* delay. We have designed and custom-built a pair of fiber collimators for fine focal adjustment, coupling, and alignment. The double-pass delay line insertion loss is less than $0.4 \pm 0.05$ dB over the entire 27 mm delay-travel range (of up to 360 ps in the reflected double-pass configuration). To verify the Franson interferometer stability, *arm1* and *arm2* are connected in series and the interference visibility is observed up to $49.8 \pm 1.0$ % for both short and long-term, near or right at the 50 % classical limit. The delay-temperature tuning transduction is quantified at 127 attoseconds per mK. More details on the characterization, setup, and design are illustrated in the Supplementary Information.

**Acknowledgements**

We acknowledge assistance from Todd Pittman, James Franson, and Dov Fields. We also acknowledge discussions with Abhinav Kumar Vinod, Yongnan Li, Joseph Poekert, Mark Itzler, Peizhe Li, Dirk R. Englund, and Xiaolong Hu. This work is supported by the InPho program from Defense Advanced Research Projects Agency (DARPA) under contract number W911NF-10-1-0416. Y.X.G is supported by the National Natural Science Foundations of China (Grant No. 11474050).

**Author contributions**

Z.X, T.Z., S.S., X.X., and J.L. performed the measurements, J.C.B. and A.R. developed the 1.3 GHz detectors, T.Z., Y.X.G., F.N.C.W., and J.H.S. provided the theory and samples, and all authors helped in the manuscript preparation.

**Additional information**

The authors declare no competing financial interests. Supplementary information accompanies this paper online. Reprints and permission information is available online at http://www.nature.com/reprints/. Correspondence and requests for materials should be addressed to Z.X. (zhenda@seas.ucla.edu) and C.W.W. (cheewei.wong@ucla.edu).

# Figures

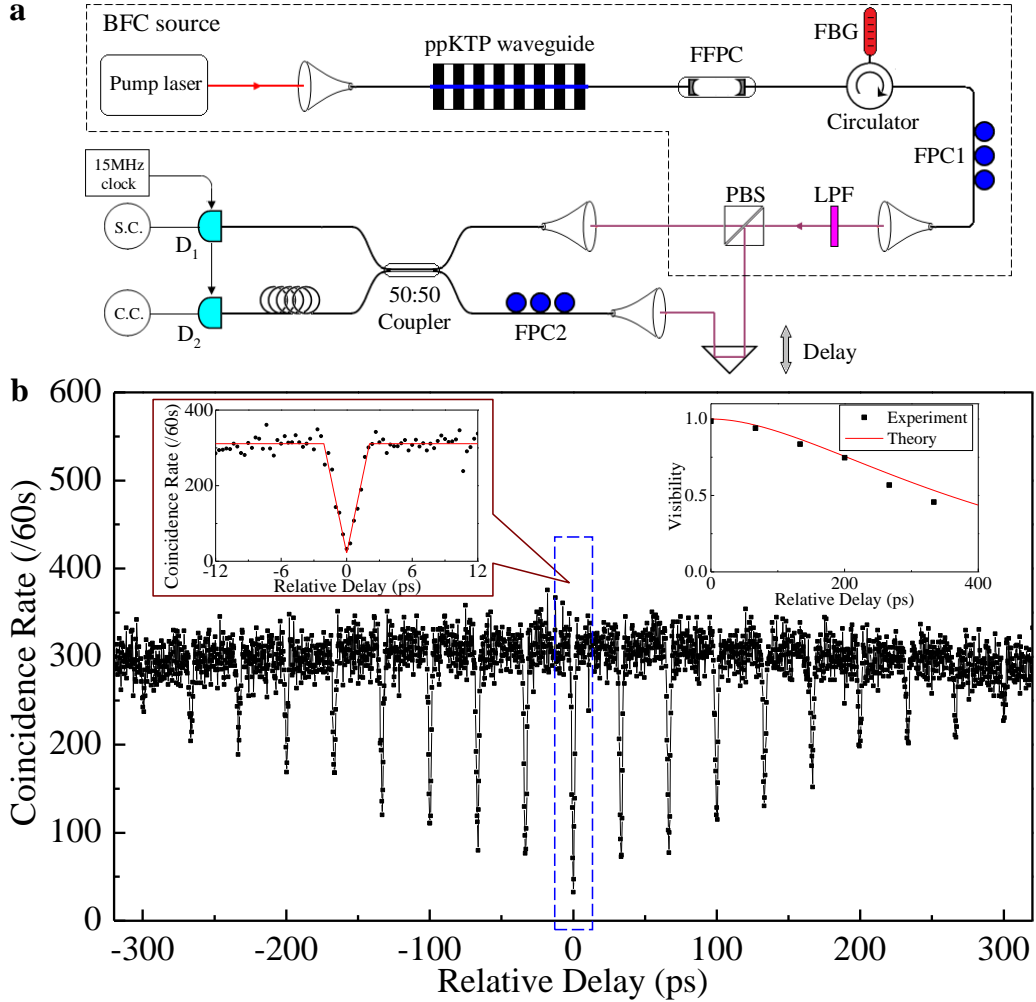

**Figure 1 | Generation and quantum revival observations of the high-dimensional biphoton frequency comb. a,** Illustrative experimental scheme. FFPC: fiber Fabry-Perot cavity; FBG: fiber Bragg grating; FPC: fiber polarization controller; LPF: long pass filter; PBS: polarizing beamsplitter; P: polarizer; S.C.: single counts; C.C.: coincidence counts. **b,** Coincidence counting rate as a function of the relative delay $\Delta T$ between the two arms of the HOM interferometer. The HOM revival is observed in the two-photon interference, with dips at 19 time-bins in this case. The visibility change across the different relative delays arises from the single FFPC bandwidth $\Delta\omega$. The red solid line is the theoretical prediction from the phase-matching bandwidth. Left insert: zoom-in coincidence around zero relative delay between the two arms. The dip width was fit to be

3.86 ± 0.30 ps, which matches well with the 245 GHz phase-matching bandwidth. The measured visibility of the dip is observed at 87.2 ± 1.5 %, or 96.5 % after subtracting the accidental coincidence counts. Right insert: measured bin visibility versus HOM delay, compared with theoretical predictions (Supplementary Information I).

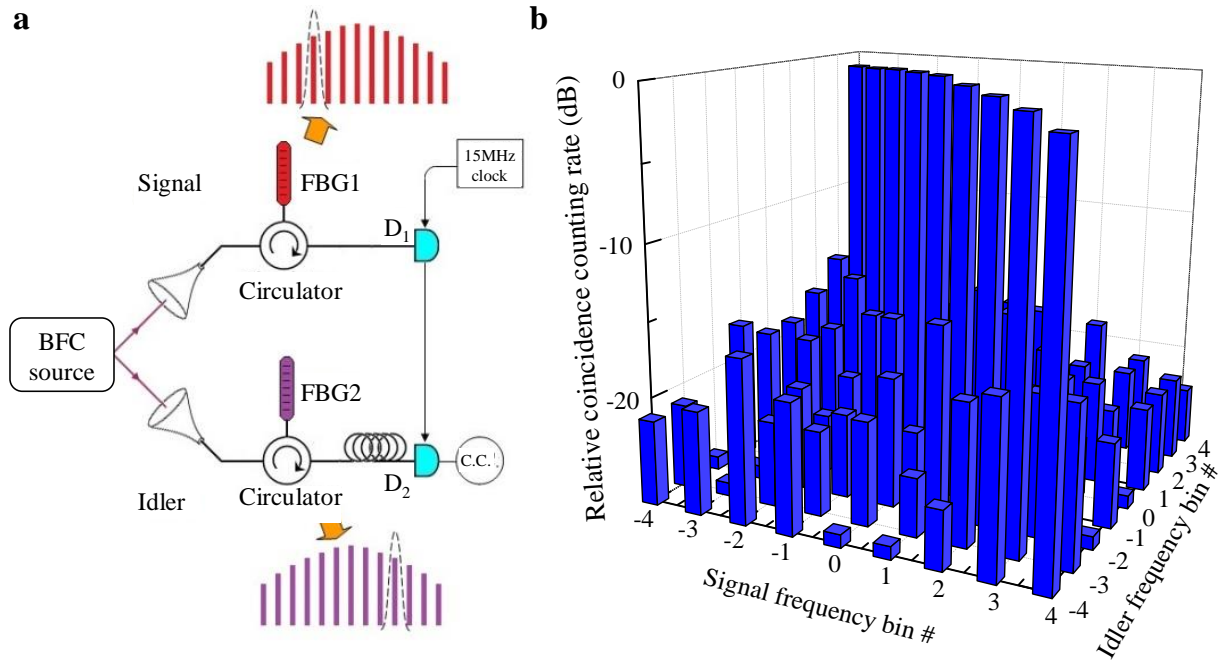

**Figure 2** | **Quantum frequency correlation measurement of the biphoton frequency comb. a,** Experimental schematic for the frequency correlation measurement. Signal and idler photons are sent to two narrow band filters for the frequency bin correlation measurement with coincidence counting. Each filter consists of a FBG and a circulator. The FBGs have matched FWHM bandwidth of 100 pm and are thermally tuned for the scans from -4th to +4th frequency bins from the center. **b,** Measured frequency correlation of the BFC. The relative coincidence counting rate is recorded while the signal and idler filters are set at different frequency-bin numbers.

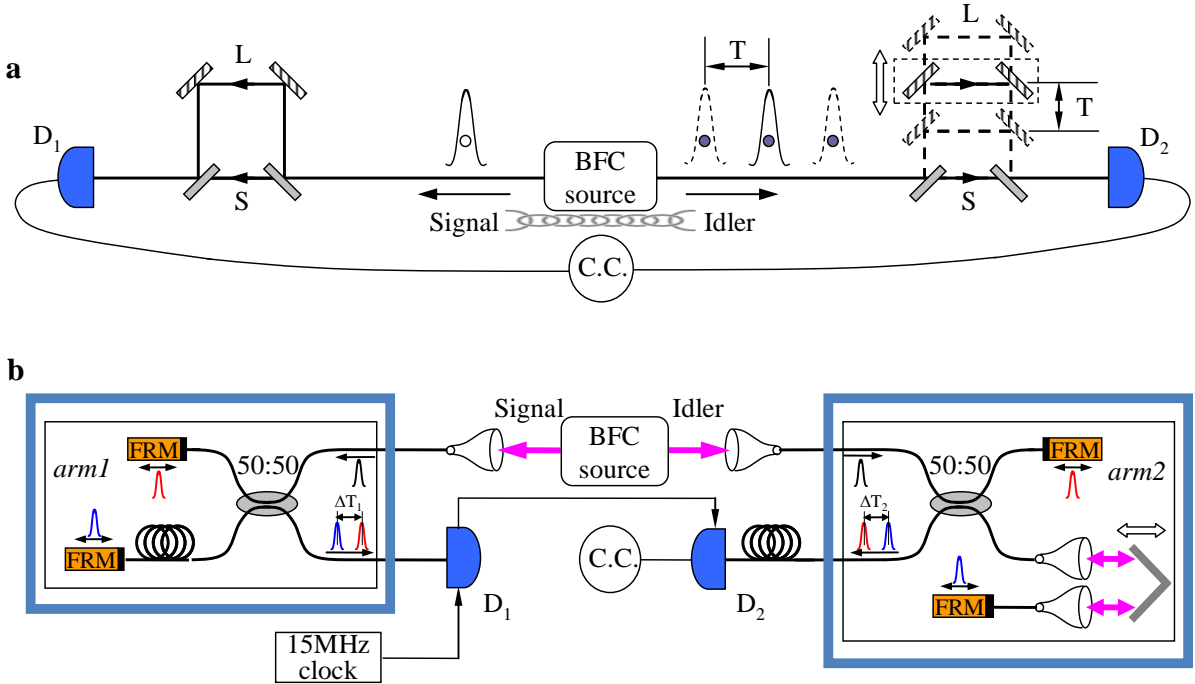

**Figure 3 | Franson interference of the high-dimensional biphoton frequency comb. a,** Schematic map and concept of Franson interference of the BFC. The BFC is prepared with high-dimensional correlation features of mode-locked behavior with repetition period of *T*. Therefore, Franson-type interference between the long-long and short-short events can be observed when $\Delta T = NT$ where *N* is an integer. **b,** Experimental Franson interference setup. FRM: Faraday mirror. The FRMs are used to compensate the stress-induced birefringence of the single-mode fiber interferometers. A compact optical delay line was used in the longer path of the *arm2* to achieve different imbalances $\Delta T$ ($=\Delta T_2 - \Delta T_1$). Both arms are double-temperature stabilized, first on the custom aluminum plate mountings and second on the sealed enclosures (light blue thick lines).

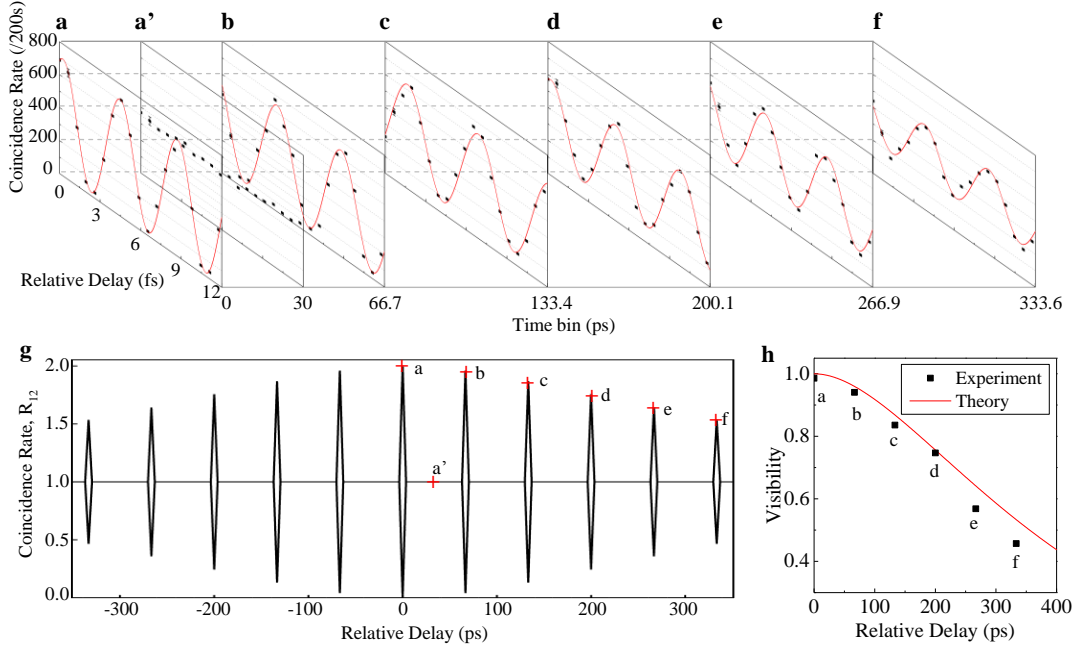

**Figure 4 | Measured Franson interference around different relative delays of *arm2*. a** to **f,** Franson interferences at time bin #0 ($\Delta T = 0$), #1 ($\Delta T = 66.7$ ps), #2 ($\Delta T = 133.4$ ps), #3 ($\Delta T = 200.1$ ps), #4 ($\Delta T = 266.9$ ps), and #5 ($\Delta T = 333.6$ ps) respectively. Also included in panel **a'** is the interference measured *away* from the above time bins at $\Delta T = 30$ ps, with no observable interference fringes. The data points in each panel (of the different relative Franson delays) include the measured error bars across each data set, arising from Poisson statistics, experimental drift, and measurement noise. The error bars from repeated coincidence measurements are much smaller than the observed coincidence rates in our measurement and setup. In each panel, the red line denotes the numerical modeling of the Franson interference on the high-dimensional quantum state. **g,** Theoretical fringe envelope of Franson interference for the high-dimensional biphoton frequency comb, with superimposed experimental observations. The marked labels (**a** to **f,** and **a'**) correspond to the actual delay points from which the above measurements are taken. **h**, Witnessed visibility of high-dimensional Franson interference fringes as a function of $\Delta T$. The experimental (and theoretical) witnessed visibilities for the *k*-th order peaks are 97.8 % (100 %), 93.3 % (96.0 %), 83.0 % (86.8 %), 74.1 % (75.6 %), 59.0 % (64.0 %), and 45.4 % (53.3 %) respectively.

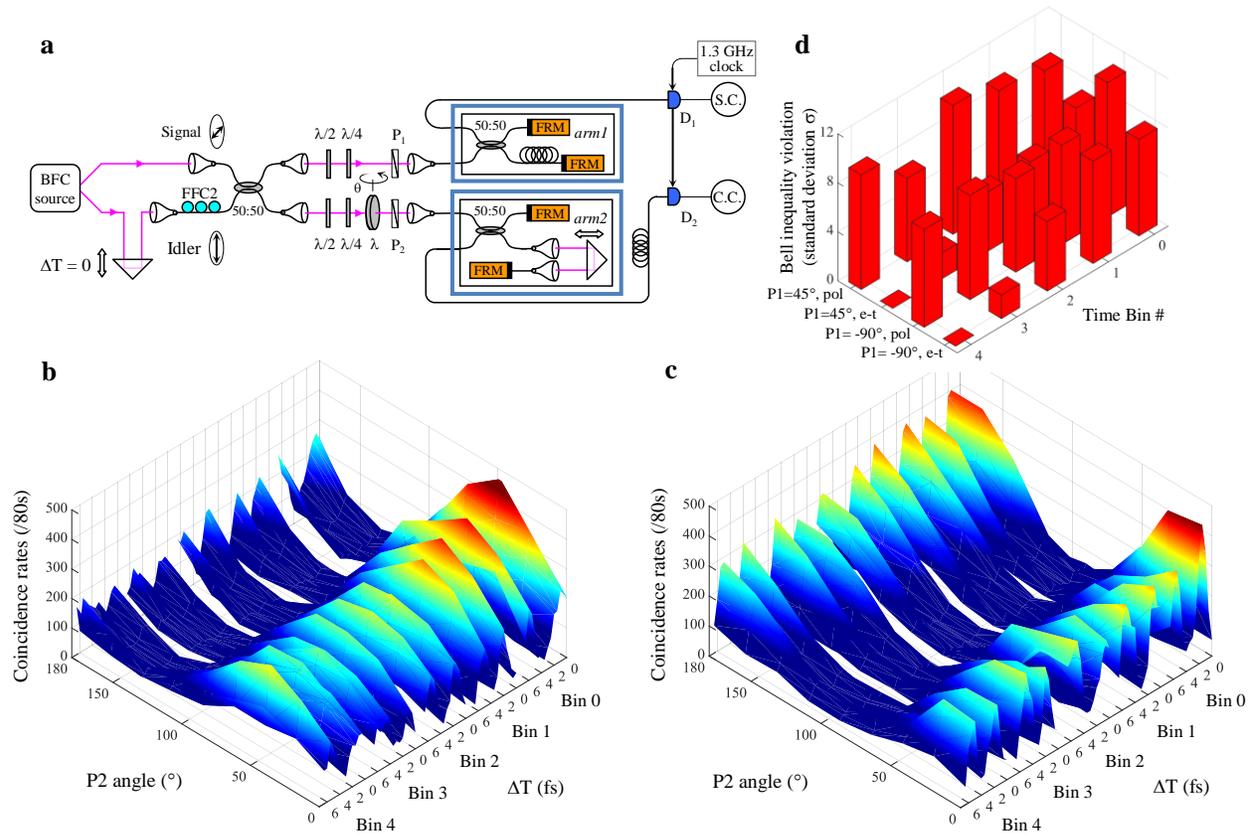

**Figure 5 | High-dimensional hyperentanglement on polarization and energy-time basis. a,** Setup for the high-dimensional two degree-of-freedom entanglement measurement. The state is generated by mixing the signal and idler photons at the 50:50 fiber beam splitter with orthogonal polarizations. Perfect temporal overlap between signal and idler photons are ensured by the HOM interference discussed before. High-dimensional hyperentanglement is measured with polarization analysis using polarizers P1 and P2, cascaded with a Franson interferometer. $\lambda/2$: half wave plate; $\lambda/4$: quarter wave plate; $\lambda$: multi-order full wave plate. **b** and **c**: measured two-photon interference fringes when P1 is set at 45 and 90 degrees. The period variance of the fringe on the temporal domain is because of a slow pump laser drift. **d,** Measured Bell inequality violation at different time-bins and P1 angles. pol: polarization basis, e-t: energy-time basis.

# Supplementary Information for

# Harnessing high-dimensional hyperentanglement through a biphoton frequency comb

Zhenda Xie[1,2], Tian Zhong[3], Sajan Shrestha[2], XinAn Xu[2], Junlin Liang[2], Yan-Xiao Gong[4], Joshua C. Bienfang[5], Alessandro Restelli[5], Jeffrey H. Shapiro[3], Franco N. C. Wong[3], and Chee Wei Wong[1,2]

[1] *Mesoscopic Optics and Quantum Electronics Laboratory, University of California, Los Angeles, CA 90095*

[2] *Optical Nanostructures Laboratory, Columbia University, New York, NY 10027*

[3] *Research Laboratory of Electronics, Massachusetts Institute of Technology, Cambridge, MA 02139*

[4] *Department of Physics, Southeast University, Nanjing, 211189, People's Republic of China*

[5] *Joint Quantum Institute, University of Maryland and National Institute of Standards and Technology, Gaithersburg, Maryland 20899, USA*

## I. Theory of two-photon interference of the high-dimensional biphoton frequency comb

Considering the Hong-Ou-Mandel (HOM) interference at an ideal 50:50 coupler, we can write the electric field operators at the two detectors $D_1$ and $D_2$ as

$$\hat{E}_1(t) = \frac{1}{\sqrt{2}}\left[\hat{E}_s(t) + \hat{E}_i(t+\delta T)\right], \quad \hat{E}_2(t) = \frac{1}{\sqrt{2}}\left[\hat{E}_s(t) - \hat{E}_i(t+\delta T)\right], \quad (1)$$

with the field operators before the HOM interferometer $\hat{E}_k(t)$, ($k=s,i$) given by

$$\hat{E}_k(t) = \frac{1}{\sqrt{2\pi}}\int d\omega\, \hat{a}_k(\omega) e^{-i\omega t}, \quad (2)$$

where $\delta T$ is the arrival time difference for the signal and idler photons from the crystal to the coupler. Then the two-photon coincidence detection rate is expressed as

$$R_{12} \propto \int_{T_g} d\tau\, G_{12}^{(2)}(t, t+\tau), \tag{3}$$

with the correlation function given by

$$G_{12}^{(2)}(t,t+\tau) = \langle \psi | \hat{E}_1^\dagger(t)\hat{E}_2^\dagger(t+\tau)\hat{E}_2(t+\tau)\hat{E}_1(t) | \psi \rangle = \left| \langle 0 | \hat{E}_2(t+\tau)\hat{E}_1(t) | \psi \rangle \right|^2, \tag{4}$$

where $T_g$ represents the timing between the detection gates. Here we assume the pump light is an ideal continuous-wave laser and thus neglect the average over the pump field. Substituting Eqs. (1), (2) and the spontaneous parametric downconverted (SPDC) state into Eq. (4) we obtain

$$G_{12}^{(2)}(t,t+\tau) \propto \left| g(\tau+\delta T) - g(-\tau+\delta T) \right|^2, \tag{5}$$

where we define $g(t) \equiv \int \Phi(\Omega) e^{i\Omega t} d\Omega$ and $\Phi(\Omega)$ denotes the spectrum amplitude. As $T_g \gg T_c$, where $T_c$ is the biphoton correlation time. The time integral range in Eq. (3) can be extended to $(-\infty, +\infty)$. Then after the time integral we obtain the coincidence rate

$$\begin{aligned} R_{12} &\propto 1 - \text{Re}\left[ \int g^*(\tau) g(\tau+\delta T) d\tau \right] \Big/ \int |g(\tau)|^2 d\tau \\ &\propto 1 - \text{Re}\left[ \int \Phi(-\Omega)\Phi(\Omega) e^{2i\Omega\delta T} d\Omega \right] \Big/ \int |\Phi(\Omega)|^2 d\Omega \end{aligned} \tag{6}$$

For our source, the spectrum amplitude $\Phi(\Omega)$ has the following form

$$\Phi(\Omega) = \sum_{m=-N}^{N} f'(\Omega) h(\Omega) f(\Omega - m\Delta\Omega) = \sum_{m=-N}^{N} \frac{\text{rect}(\Omega/B)\,\text{sinc}(A\Omega)}{(\Delta\omega)^2 + (\Omega - m\Delta\Omega)^2}, \tag{7}$$

where $f'(\Omega) = \text{sinc}(A\Omega)$ is the phase matching spectrum function, with the full width at half maximum (FWHM) as $2.78/A$, and $h(\Omega) = \text{rect}(\Omega/B) = \begin{cases} 1, & |\Omega/B| \leq 1/2 \\ 0, & |\Omega/B| > 1/2 \end{cases}$ is the rectangular spectrum function resulting from the FBG, with $B$ denoting the width. The fiber Fabry-Perot cavity (FFPC) has a Lorentzian spectrum bin function characterized by: $f(\Omega) = f_s(\Omega) f_i(\Omega) = 1/[(\Delta\omega + i\Omega)(\Delta\omega - i\Omega)] = 1/[(\Delta\omega)^2 + \Omega^2]$, where $\Delta\Omega$ is the spacing between the frequency bins and $2\Delta\omega$ denotes the FWHM of each frequency bin. Corresponding to our measurements, we evaluate our theory using the following parameter values: $A = 2.78/(2\pi \times$

245 GHz) = 1.81 ps, $B = 2\pi \times 346$ GHz $= 2.2 \times 10^{12}$ rad/s, $\Delta\Omega = 2\pi \times 15.15$ GHz $= 95.2 \times 10^{10}$ rad/s, and $\Delta\omega = 2\pi \times 1.36$ GHz $/ 2 = 4.27 \times 10^9$ rad/s. The spectrum density $|\Phi(\Omega)|^2$ can be described by

$$|\Phi(\Omega)|^2 = \sum_{m=-N}^{N} |f'(\Omega)|^2 |h(\Omega)|^2 |f(\Omega - m\Delta\Omega)|^2 = \sum_{m=-N}^{N} \frac{\text{sinc}^2(A\Omega)\text{rect}(\Omega/B)}{[(\Delta\omega)^2 + (\Omega - m\Delta\Omega)^2]^2}, \quad (8)$$

where we have neglected the overlaps among the frequency bins since $\Delta\Omega \gg 2\Delta\omega$. Moreover, as the FWHM of $|f'(\Omega)|^2$, $2.78/A$ is much larger than $2\Delta\omega$ and the width of $|h(\Omega)|^2$, and $B \gg 2\Delta\omega$, we can make the following approximation

$$|\Phi(\Omega)|^2 \cong \sum_{m=-N}^{N} |f'(Am\Delta\Omega)|^2 |h(m\Delta\Omega)|^2 |f(\Omega - m\Delta\Omega)|^2$$
$$= \sum_{m=-N}^{N} \frac{\text{sinc}^2(Am\Delta\Omega)\text{rect}(m\Delta\Omega/B)}{[(\Delta\omega)^2 + (\Omega - m\Delta\Omega)^2]^2} = \sum_{m=-N_0}^{N_0} \frac{\text{sinc}^2(Am\Delta\Omega)}{[(\Delta\omega)^2 + (\Omega - m\Delta\Omega)^2]^2}, \quad (9)$$

where $N_0 = \lfloor B/(2\Delta\Omega) \rfloor = 11$ is the integer part of $B/2\Delta\Omega$. Thus we can obtain

$$\int |\Phi(\Omega)|^2 d\Omega = \sum_{m=-N_0}^{N_0} \int d\Omega \frac{\text{sinc}^2(Am\Delta\Omega)}{[(\Delta\omega)^2 + (\Omega - m\Delta\Omega)^2]^2} = \frac{\pi}{2(\Delta\omega)^3} \sum_{m=-N_0}^{N_0} \text{sinc}^2(Am\Delta\Omega). \quad (10)$$

$$\int \Phi(-\Omega)\Phi(\Omega)e^{2i\Omega\delta T} d\Omega = \int |\Phi(\Omega)|^2 e^{2i\Omega\delta T} d\Omega$$
$$= \sum_{m=-N_0}^{N_0} \int d\Omega \frac{\text{sinc}^2(Am\Delta\Omega)}{[(\Delta\omega)^2 + (\Omega - m\Delta\Omega)^2]^2} e^{2i\Omega\delta T}$$
$$= \frac{\pi e^{-2\Delta\omega|\delta T|}(1 + 2\Delta\omega|\delta T|)}{2(\Delta\omega)^3} \sum_{m=-N_0}^{N_0} \text{sinc}^2(Am\Delta\Omega)e^{2im\Delta\Omega\delta T} \quad (11)$$
$$= \frac{\pi e^{-2\Delta\omega|\delta T|}(1 + 2\Delta\omega|\delta T|)}{2(\Delta\omega)^3} \left[ 2\sum_{m=1}^{N_0} \text{sinc}^2(Am\Delta\Omega)\cos(2m\Delta\Omega\delta T) + 1 \right].$$

Then through Eq. (6) we arrive at the coincidence rate as

$$R_{12} \propto 1 - \frac{e^{-2\Delta\omega|\Delta T|}(1 + 2\Delta\omega|\delta T|)}{\sum_{m=-N_0}^{N_0} \text{sinc}^2(Am\Delta\Omega)} \left[ 2\sum_{m=1}^{N_0} \text{sinc}^2(Am\Delta\Omega)\cos(2m\Delta\Omega\delta T) + 1 \right]. \quad (12)$$

We model the second-order correlation based on our experimental parameters. The coincidence rate versus $\delta T$ in the range of {-320 ps, 320 ps} is plotted in Supplementary Figure S1. We can see the interference fringe is a multi-dip pattern, with the dip revival period of about 33.2 ps, matching the 33.4 ps in our measurements. The linewidth of each Hong-Ou-Mandel dip and the fall-off in visibility for increasing $\pm k$ bins from the zero delay point also matches with the experimental observations.

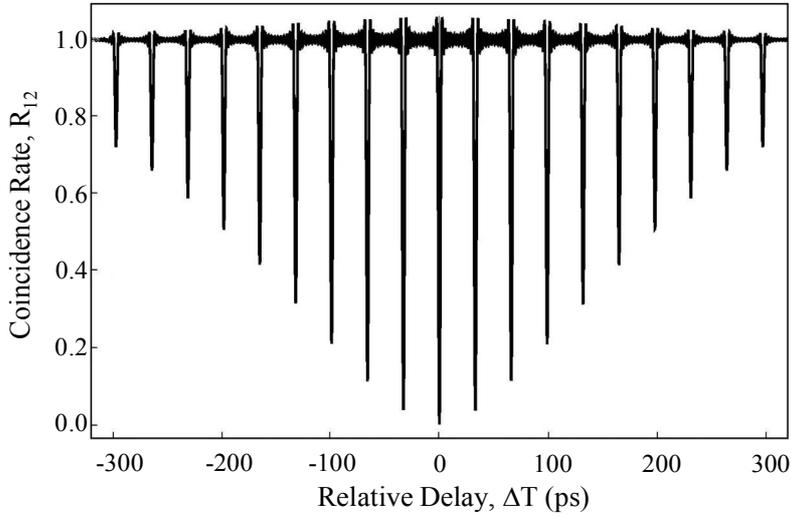

**Supplementary Figure S1 | Modeling of Hong-Ou-Mandel interference revivals for the high-dimensional biphoton frequency comb.** The coincidence counting rate $R_{12}$ as a function of $\delta T$, the arrival time difference between the signal and idler photons. The fall-off in the revived visibility away from the zero delay point arises from the Lorentzian lineshape of the SPDC individual photons after passing through the cavity.

In the experiment, the bandwidth of FBG is more than that of the phase matching. It is reasonable to assume an infinite number of frequency bins, i.e., $N \to \infty$, and making the replacement of $\delta T = \delta T' + kT/2$, where $T = 2\pi/\Delta\Omega$, $-A < \delta T' \leq T/2 - A$, and $k$ is an integer number, we can simplify Eq. (12) to

$$R_{12}(\delta T'+kT/2) \propto$$
$$\begin{cases} 1-e^{-2\Delta\omega|\delta T'+k\pi/\Delta\Omega|}(1+2\Delta\omega|\delta T'+k\pi/\Delta\Omega|)(1-|\delta T'|/A) & -A < \delta T' < A, \\ 0 & A \leq \delta T' \leq \pi/\Delta\Omega - A. \end{cases} \quad (13)$$

We can see the dip revival period is $T_R = \pi/\Delta\Omega \approx 33.2$ ps (or $T/2$) and the visibility of the $k$-th order dip is $e^{-2|k|\pi\Delta\omega/\Delta\Omega}(1+2|k|\pi\Delta\omega/\Delta\Omega)$. We note that the recurrence is at $T/2$ instead of at $T$ in our prior analysis where the beamsplitter was moved [J. H. Shapiro, *Technical Digest of Topical Conference on Nonlinear Optics*, p.44, FC7-1, Optical Society of America (2002)]. With the beamsplitter shift, the transmitted signal and idler beams do not experience an advance or a delay, but one reflected beam is advanced while the other is delayed. In our case, only the idler beam is delayed, giving rise to the *T/2* recurrence in the coincidences as detailed above.

## II. Theory of Franson interference of the high-dimensional biphoton frequency comb

The electric field operators at the two detectors $D_1$ and $D_2$ can be expressed as

$$\hat{E}_1(t) = \frac{1}{2}\left[\hat{E}_s(t) + \hat{E}_s(t-\Delta T_1)\right], \quad \hat{E}_2(t) = \frac{1}{2}\left[\hat{E}_i(t) + \hat{E}_i(t-\Delta T_2)\right], \quad (14)$$

with the field detectors before the Franson interferometer $\hat{E}_k(t)$, ($k=s,i$) given by

$$\hat{E}_k(t) = \frac{1}{\sqrt{2\pi}} \int d\omega \hat{a}_k(\omega) e^{-i\omega t}, \quad (15)$$

where $\Delta T_1, \Delta T_2$ are the unbalanced arm differences. Then the two-photon coincidence detection rate can be described by

$$R_{12} \propto \int_{T_g} d\tau G_{12}^{(2)}(t, t+\tau), \quad (16)$$

with the correlation function given by

$$G_{12}^{(2)}(t, t+\tau) = \langle \psi | \hat{E}_1^\dagger(t) \hat{E}_2^\dagger(t+\tau) \hat{E}_2(t+\tau) \hat{E}_1(t) | \psi \rangle = \left|\langle 0 | \hat{E}_2(t+\tau) \hat{E}_1(t) | \psi \rangle\right|^2, \quad (17)$$

where $T_g$ represent the timing between the detection gates. Here we assume the pump light is an ideal continuous-wave laser and thus neglect the average over the pump field. Substituting Eqs. (1), (2) and the SPDC state into Eq. (4), we obtain

$$G^{(2)}_{12}(t,t+\tau) = |G(t,t) + G(t-\Delta T_1, t-\Delta T_2) + G(t-\Delta T_1, t) + G(t, t-\Delta T_2)|^2, \tag{18}$$

with

$$G(t_1,t_2) = \frac{1}{8\pi} e^{-i\omega_p(t_1+t_2+\tau)/2} \int \Phi(\Omega) e^{i\Omega(t_2-t_1+\tau)} d\Omega = \frac{1}{8\pi} e^{-i\omega_p(t_1+t_2+\tau)/2} g(t_2-t_1+\tau), \tag{19}$$

where $\Phi(\Omega)$ denotes the spectrum amplitude and we define $g(t) \equiv \int \Phi(\Omega) e^{i\Omega t} d\Omega$. Then we may rewrite Eq. (5) as

$$\begin{aligned}G^{(2)}_{12}(t,t+\tau) \propto \big| & e^{-i\omega_p\tau/2} g(\tau) + e^{-i\omega_p(\tau-\Delta T_1-\Delta T_2)/2} g(\tau+\Delta T_1-\Delta T_2) \\ & + e^{-i\omega_p(\tau-\Delta T_1)/2} g(\tau+\Delta T_1) + e^{-i\omega_p(\tau-\Delta T_2)/2} g(\tau-\Delta T_2) \big|^2.\end{aligned} \tag{20}$$

As we have noted, $\Delta T_1, \Delta T_2$ are much larger than the single-photon coherence time $T_c$, i.e., the range of the function $g(t)$, so there is only one non-zero cross term in Eq. (20). Thus we obtain

$$\begin{aligned}G^{(2)}_{12}(t,t+\tau) \propto & |g(\tau)|^2 + |g(\tau+\Delta T_1-\Delta T_2)|^2 + |g(\tau+\Delta T_1)|^2 + |g(\tau-\Delta T_2)|^2 \\ & + 2\text{Re}[e^{i\omega_p(\Delta T_1+\Delta T_2)/2} g^*(\tau) g(\tau+\Delta T_1-\Delta T_2)].\end{aligned} \tag{21}$$

Since $\Delta T_1, \Delta T_2 \gg T_g$, the coincidence detection system can resolve the short and long paths and thus the two terms $|g(\tau+\Delta T_1)|^2$, $|g(\tau-\Delta T_2)|^2$ have no contribution to the coincidence rate. Moreover, as $T_g \gg T_c$, the time integral range in Eq. (3) can be extended as $(-\infty, +\infty)$. Then after the time integral we obtain the coincidence rate

$$R_{12} \propto 1 + |\Gamma(\Delta T)| \cos[[\omega_p \Delta T/2 + \omega_p \Delta T_2 + \varphi], \tag{22}$$

where $\Delta T = \Delta T_1 - \Delta T_2$, and

$$\begin{aligned}\Gamma(\Delta T) &= \int g^*(\tau) g(\tau+\Delta T) d\tau \Big/ \int |g(\tau)|^2 d\tau \\ &= \int \Phi(-\Omega)\Phi(\Omega) e^{i\Omega\Delta T} d\Omega \Big/ \int |\Phi(\Omega)|^2 d\Omega = |\Gamma(\Delta T)| e^{i\varphi}.\end{aligned} \tag{23}$$

For our source, the spectrum amplitude $\Phi(\Omega)$ has the following form

$$\Phi(\Omega) = \sum_{m=-N}^{N} f'(\Omega)h(\Omega)f(\Omega - m\Delta\Omega) = \sum_{m=-N}^{N} \frac{\text{rect}(\Omega/B)\text{sinc}(A\Omega)}{(\Delta\omega)^2 + (\Omega - m\Delta\Omega)^2}, \quad (24)$$

Then we can write the spectrum density $|\Phi(\Omega)|^2$ as

$$|\Phi(\Omega)|^2 = \sum_{m=-N}^{N} |f'(\Omega)|^2 |h(\Omega)|^2 |f(\Omega - m\Delta\Omega)|^2 = \sum_{m=-N}^{N} \frac{\text{sinc}^2(A\Omega)\text{rect}(\Omega/B)}{[(\Delta\omega)^2 + (\Omega - m\Delta\Omega)^2]^2}, \quad (25)$$

where we have neglected the overlaps among the frequency bins since $\Delta\Omega \gg 2\Delta\omega$. Moreover, as the FWHM of $|f'(\Omega)|^2$, $2.78/A \gg 2\Delta\omega$, and the width of $|h(\Omega)|^2$, $B \gg 2\Delta\omega$, we can make the following approximation

$$|\Phi(\Omega)|^2 \cong \sum_{m=-N}^{N} |f'(Am\Delta\Omega)|^2 |h(m\Delta\Omega)|^2 |f(\Omega - m\Delta\Omega)|^2$$
$$= \sum_{m=-N}^{N} \frac{\text{sinc}^2(Am\Delta\Omega)\text{rect}(m\Delta\Omega/B)}{[(\Delta\omega)^2 + (\Omega - m\Delta\Omega)^2]^2} = \sum_{m=-N_0}^{N_0} \frac{\text{sinc}^2(Am\Delta\Omega)}{[(\Delta\omega)^2 + (\Omega - m\Delta\Omega)^2]^2}, \quad (26)$$

where $N_0 = \lfloor B/(2\Delta\Omega) \rfloor = 11$ is the integer part of $B/(2\Delta\Omega)$. Thus we can obtain

$$\int |\Phi(\Omega)|^2 d\Omega = \sum_{m=-N_0}^{N_0} \int d\Omega \frac{\text{sinc}^2(Am\Delta\Omega)}{[(\Delta\omega)^2 + (\Omega - m\Delta\Omega)^2]^2} = \frac{\pi}{2(\Delta\omega)^3} \sum_{m=-N_0}^{N_0} \text{sinc}^2(Am\Delta\Omega). \quad (27)$$

$$\int \Phi(-\Omega)\Phi(\Omega)e^{i\Omega\Delta T} d\Omega = \int |\Phi(\Omega)|^2 e^{i\Omega\Delta T} d\Omega$$
$$= \sum_{m=-N_0}^{N_0} \int d\Omega \frac{\text{sinc}^2(Am\Delta\Omega)}{[(\Delta\omega)^2 + (\Omega - m\Delta\Omega)^2]^2} e^{i\Omega\Delta T}$$
$$= \frac{\pi e^{-\Delta\omega|\Delta T|}(1+\Delta\omega|\Delta T|)}{2(\Delta\omega)^3} \sum_{m=-N_0}^{N_0} \text{sinc}^2(Am\Delta\Omega)e^{im\Delta\Omega\Delta T} \quad (28)$$
$$= \frac{\pi e^{-\Delta\omega|\Delta T|}(1+\Delta\omega|\Delta T|)}{2(\Delta\omega)^3} \left[ 2\sum_{m=1}^{N_0} \text{sinc}^2(Am\Delta\Omega)\cos(m\Delta\Omega\Delta T) + 1 \right].$$

Then through Eqs. (6) and (10), we can write the coincidence rate as

$$R_{12} \propto 1 + \frac{e^{-\Delta\omega|\Delta T|}(1+\Delta\omega|\Delta T|)}{\sum_{m=-N_0}^{N_0} \text{sinc}^2(Am\Delta\Omega)} \left[ 2\sum_{m=1}^{N_0} \text{sinc}^2(Am\Delta\Omega)\cos(m\Delta\Omega\Delta T) + 1 \right] \quad (29)$$
$$\times \cos[\omega_p(\Delta T/2 + \Delta T_2)].$$

Since $\omega_p/2 = \pi c/\lambda_p = 1.43 \times 10^{12}$ rad/s $\gg \Delta\omega, \Delta\Omega$, the term $\cos[\omega_p(\Delta T/2 + \Delta T_2)]$ is the fast collision part of the interference fringe with the other part determining the fringe envelope. We can simulate our experimental results with the theoretical parameters above and for $\Delta T_2 = 5$ ns.

If we consider a large number of frequency bins such as in our measurements, i.e., $N \to \infty$, and make the replacement of $\Delta T = \Delta T' + kT$, where $-2A < \Delta T' \leq T - 2A$, and $k$ is an integer number, we can simplify Eq. (16) to

$$R_{12}(\Delta T' + kT) \propto \begin{cases} 1 + e^{-\Delta\omega|\Delta T' + 2k\pi/\Delta\Omega|}(1 + \Delta\omega|\Delta T' + 2k\pi/\Delta\Omega|)[1 - |\Delta T'|/(2A)] \\ \quad \times \cos[\omega_p(\Delta T'/2 + k\pi/\Delta\Omega + \Delta T_2)] \qquad -2A < \Delta T' < 2A, \\ 0 \qquad\qquad\qquad\qquad\qquad\qquad\qquad\qquad 2A \leq \Delta T' \leq 2\pi/\Delta\Omega - 2A. \end{cases} \quad (30)$$

The fringe envelope of the coincidence rate versus $\Delta T$ is plotted in Supplementary Figure S2. We see that the interference fringe has a recurrent envelope that falls off away from the zero delay point due to the Lorentzian lineshape of the FFPC. The recurrence period $T$ is about 66 ps and agrees well with repetition time of the biphoton frequency comb, i.e., the round trip time of the FFPC. The maximum visibility at the $k$-th order peaks can be found to be $e^{-2|k|\pi\Delta\omega/\Delta\Omega}(1 + 2|k|\pi\Delta\omega/\Delta\Omega)$. This gives an envelope visibility which matches well with the experimental measurements.

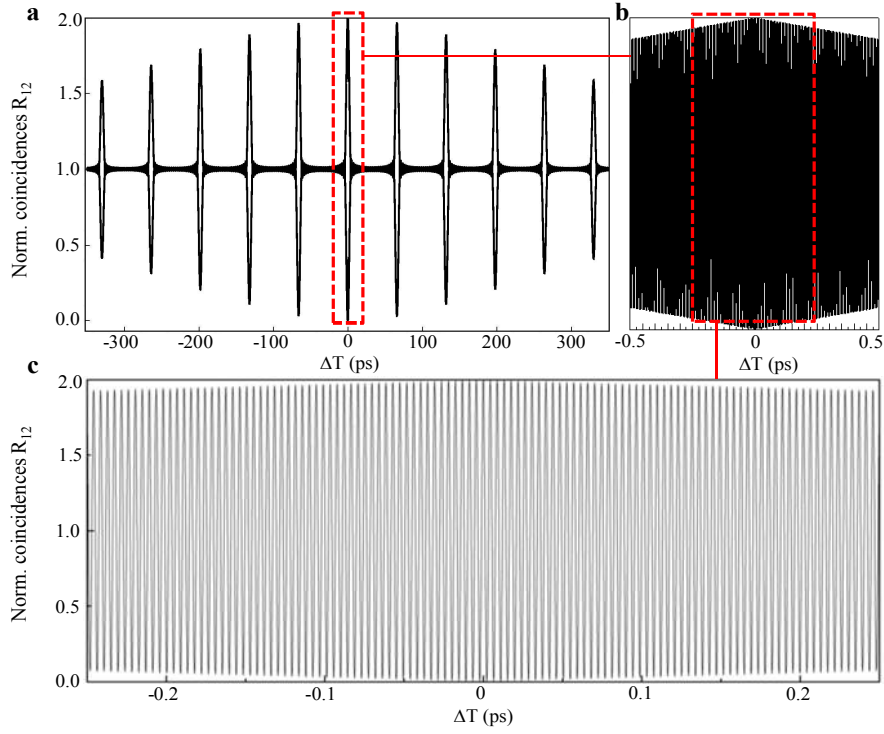

**Supplementary Figure S2 | Theory of Franson interference revival for the high-dimensional biphoton frequency comb**. **a,** The envelope of coincidence counting rates $R_{12}$ plotted as a function of $\Delta T$. The interference has a recurrent envelope that falls off away from the zero delay point, arising from the finite coherence time of the single frequency bin. **b,** zoom-in of Franson interferences for the first time bin. **c,** Further zoom-in of the Franson interferences for the first time bin. The high-frequency interference oscillations arise from the phase.

### III. Characterization of Franson interferometer long-term stability

In our measurement, the fiber-based Franson interferometer needs to be stabilized at the wavelength level over long term for the phase sensitive interference measurements. All components are fixed on the aluminum housing with thermal conductive epoxy for good thermal contact. Both interferometer arms are temperature-controlled with Peltier modules and sealed in aluminum enclosures which themselves are also temperature stabilized. The delay line in *arm2* is based on a miniaturized linear stage with closed-loop piezoelectric motor control (CONEX-AG-LS25-27P, Newport Corporation). The two fiber collimators are custom-built in-house with fine focal adjustment for optimized pair performance and epoxied on the housing. The delay line is aligned so that the double-pass insertion loss is less than 0.4 ± 0.05 dB throughout the whole travel range of 27 mm of the linear stage. The effective delay range is about 0 to 360 ps for the reflected light, considering the double-pass optical path configuration and reflector setup.

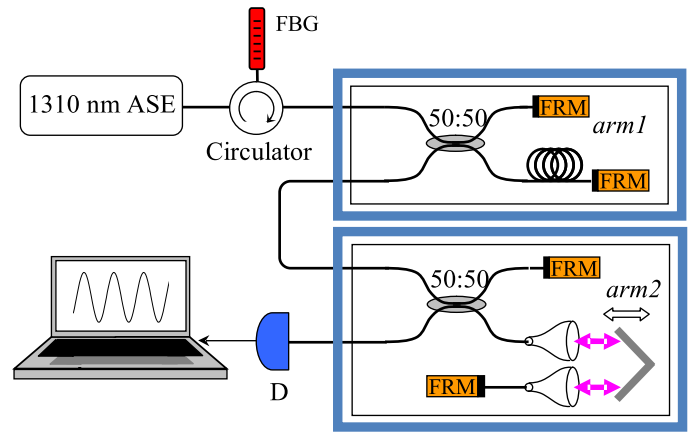

**Supplementary Figure S3 | Long-term stability test of the Franson interferometer.** *Arm1* and *arm2* are connected in series for the classical interference test. Both interferometers are double temperature controlled, and the delay line in *arm2* is closed-loop piezoelectrically controlled. Dual collimators are custom-built in-house with fine focal adjustment and optimized performance. Fine-tuned alignment is such that the double-pass insertion loss is less than 0.4 dB throughout the whole 27 mm travel range and up to 360 ps optical delay.

We have verified the stability of the Franson interferometer using classical interference before the quantum correlation measurements. The setup is shown in Supplementary Figure S3. The light is from a 1310 nm amplified spontaneous emission (ASE) source (S5FC1021S, Thorlabs Inc.), and filtered with the same filter sub-assembly that is used in the biphoton frequency comb generation. The two arms are connected in series so that classical interference occurs between the events of passing long path of *arm1*, short path of *arm2*, and short path of *arm1*, long path of *arm2*, while $\Delta T_1 = \Delta T_2$. The visibility of this interference is limited to 50%, because of other events that contribute to the background. The intensity of the output is measured using a photodiode (PDA20CS, Thorlabs Inc.), while tuning the temperature of *arm1*. The result is shown in Supplementary Figure S4. The observed interferences agree well with a sinusoidal fit, and the visibility is 49.8 ± 1.0%, near or right at the classical limit over long measurement time periods. This indicates the Franson interferometer is well stabilized. We can also obtain the delay coefficient for this temperature tuning, which is observed to be 127 attoseconds per mK based on the fringe period when the temperature is tuned.

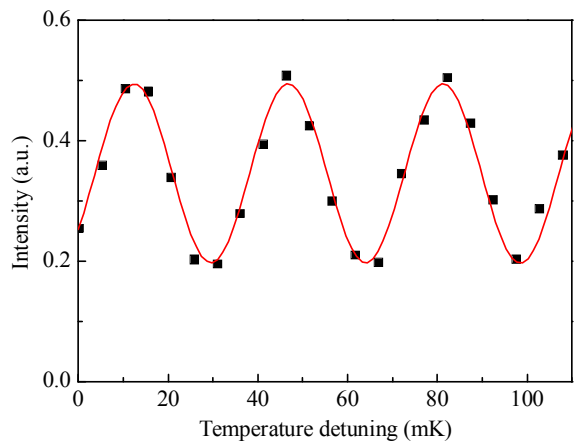

**Supplementary Figure S4 | Classical interference visibility of the stabilized coupled interferometers.** Temperature of *arm1* is tuned and the input is a 1310 nm ASE source. The temperature-delay sensitivity is observed to be 127 attoseconds per mK.

### IV. Stabilization of the pump laser in Franson interference

The Franson interference requires high stability – both short and long term – of the pump laser, which defines the two-photon coherence time of the two-photon state. The coherence time of the pump laser should be much longer than the path length difference. This path length difference is 5 ns in our experiment. Therefore, we custom-built a stabilized laser at 658 nm using self-injection-locking. The setup is shown in Supplementary Figure S5a, which is similar to a Littman–Metcalf configuration external diode laser. The laser source that we use is a standard Fabry-Perot laser diode with center wavelength around 658 nm (QLD-658-20S). It is spatially single-mode, but has multiple longitudinal modes without feedback. A diffraction grating is used for the longitudinal mode selection. The laser beam is first collimated onto the grating. The first-order diffraction reflects off a tunable mirror back into the diode through the grating. The zeroth order diffraction from the grating is used as the output. Because of the grating, we succeed in achieving single-mode lasing, with a mode-rejection ratio over 30 dB. The whole setup is isolated with double enclosures. The inner enclosure is made from aluminum and temperature-stabilized. The outer laser housing is also a solid piece of aluminum that is temperature-stabilized. A third temperature stabilization is applied to the diode.

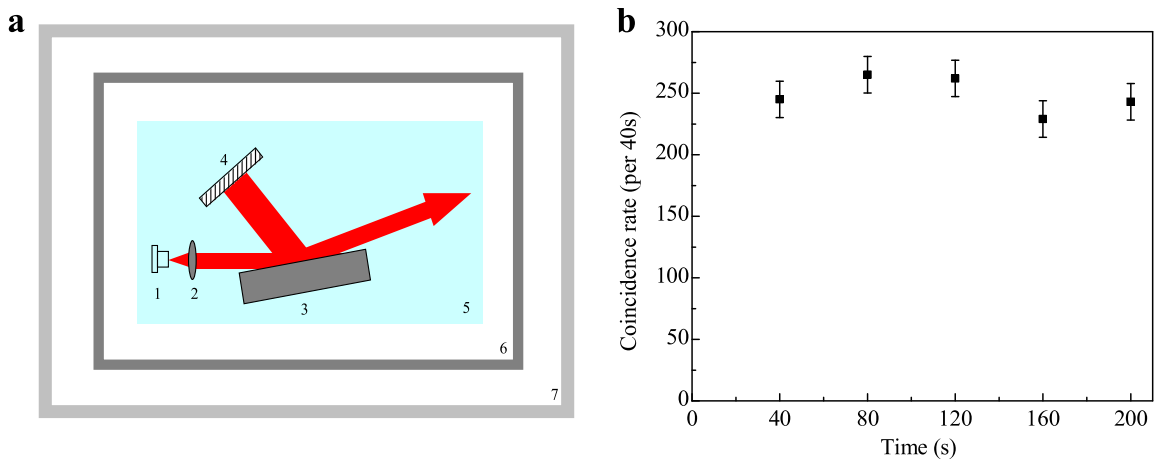

**Supplementary Figure S5 | Stabilized 658 nm pump laser with self-injection locking and characterization through Franson-type interferometer. a,** Layout schematic of custom-built 658 nm stabilized laser. 1, 658 nm Fabry-Perot laser diode with temperature stabilization; 2, collimating lens; 3, diffraction grating; 4, high-reflection mirror; 5, temperature-stabilized laser housing; 6, internal enclosure with temperature stabilization; 7, external enclosure. **b,** Franson interference measurement to demonstrate pump laser stability, with the interferometer tuned to the 0th time bin. Measurement is made at 6 mW pump power. The measured deviation of the counting rate is about 5%, with the coincidence measurements taken every 40 seconds, which corresponds to a 2 MHz drift of the pump laser. The long term drift, over 12 hours, is less than 100 MHz.

The current source for the diode is a low-noise laser diode controller (D2-105, Vescent Photonics, Inc.). With these stabilizations, we achieve a free-running wavelength drift of the 658 nm laser at less than 2 MHz within 200 seconds, which is an integration time step for the Franson measurement. The laser linewidth is measured classically with the Franson-type interferometer setup. The pump power is about 6 mW in the measurement. We tune the interferometer to the 0th time bin. $\Delta T$ is set such that the coincidence counting rate is about the middle of the sinusoidal fringe, which gives the best sensitivity to the pump drift. The coincidence counting rate is taken every 40 seconds, and the result is shown in Supplementary Figure S5b. The measured deviation of counting rate is about 5%, which corresponds to a pump drift of 2 MHz. The long-term drift is less than 100 MHz within 12 hours (measured with a wavelength meter, HighFinesse WS-7).

## V. Measurements of Hong-Ou-Mandel revival of the high-dimensional biphoton frequency comb

In the main text, we used an InGaAs/InP single-photon detector $D_2$ with ~ 2.5 ns effective gate width for the measurements, so that the detection gate widows of $D_1$ and $D_2$ are always well-overlapped through the scanning range of relative delay.

Figure S6 illustrates another example of the Hong-Ou-Mandel revival when using an effective detector gate width of ≈ 400 ps. The maximum visibility of the central dip can be enhanced to 96.1%, because of the reduced accidental coincidence possibility. That visibility becomes 97.8% after subtracting the accidental coincidence counts. We note that, in the accidentals subtraction, the estimated double pairs are still included in the counts for the best estimate of the visibility. The single-photon counting rate remains constant during the measurement. However, here the background of the coincidence counting rate drops as the relative delay increases (see Supplementary Figure S6a). This drop is from the temporal overlap reduction for the gating of $D_1$ and $D_2$ at large relative delays of the HOM interferometer.

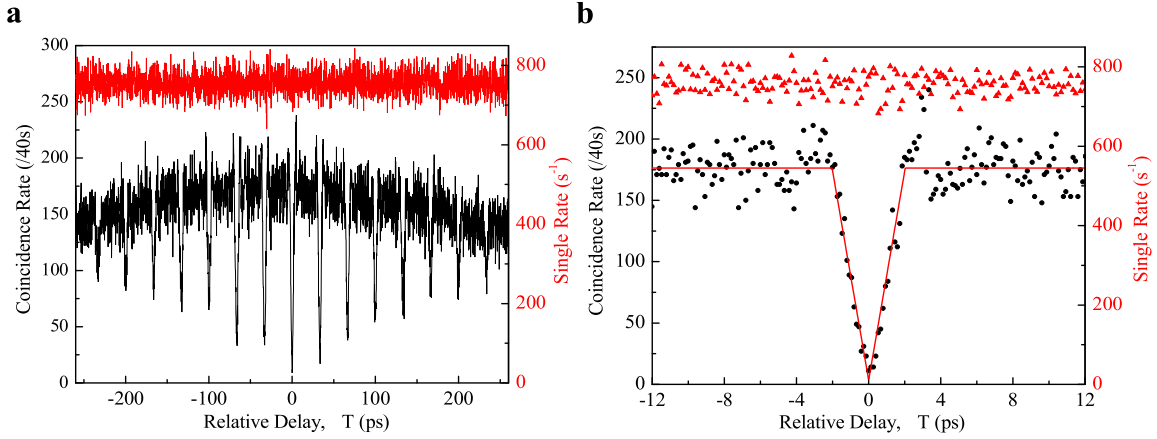

**Supplementary Figure S6 | HOM measurement with single-photon detectors with an effective gate width of ≈ 400 ps. a,** Coincidence and single counting rates as a function of the relative delay between the two arms of the HOM interferometer. The background of the coincidence counting rate drops because of the reduced overlap between the detection windows of $D_1$ and $D_2$. **b,** Zoom-in coincidence and single counting rate around zero relative delay between the two arms. The visibility is measured to be 96.1 %, or 97.8 % after subtracting the accidental coincidence counts.

## VI. Polarization entanglement measurements of the high-dimensional biphoton frequency comb

Before the hyperentanglement measurements, we test the polarization entanglement alone for the hyperentangled state. Supplementary Figure S7a shows the experiment setup. We mix the signal and idler photons on a 50:50 fiber coupler with orthogonal polarizations. By keeping the relative delay $\delta\tau = 0$, the signal and idler photons are well-overlapped temporally for the interference. We present a measurement for the polarization entanglement by measuring the coincidence counting rates while changing the angle of P2, when P1 was set at 45° and 90°, respectively. As shown in Supplementary Figure S7b, both results fit well with sinusoidal curves, with visibilities of 91.2 ± 1.6% and 93.0 ± 1.3%, which violate the Bell inequality by 12.8 and 17.7 standard deviations, respectively. This indicates the high-dimensional polarization entangled state $|\psi\rangle = \sum_{m=-N}^{N} |\omega_p/2 + m\Delta\Omega\rangle_1 |\omega_p/2 - m\Delta\Omega\rangle_2 \otimes (|H\rangle_1|V\rangle_2 + |V\rangle_1|H\rangle_2)$ is generated with high quality. Hence, in addition to the 4 frequency bits, the biphoton frequency comb also has polarization entanglement for use as a high-dimensional quantum communications platform.

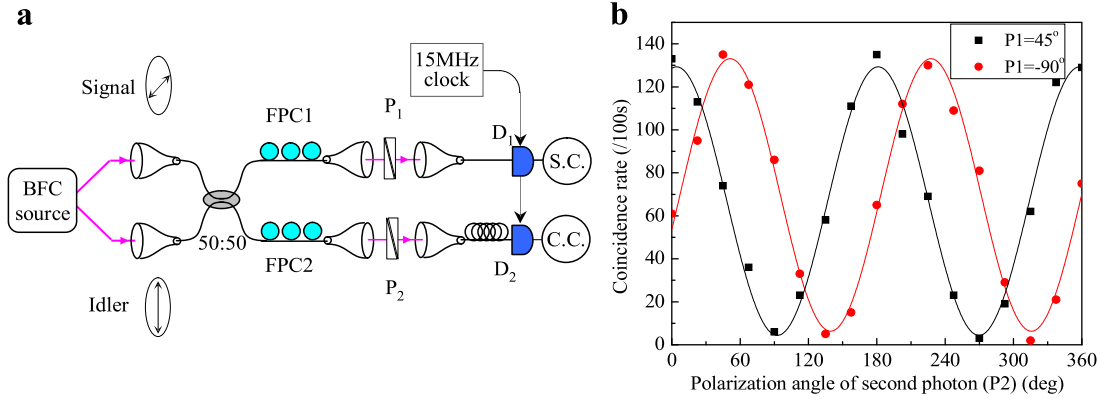

**Supplementary Figure S7 | Polarization entanglement measurements of the high-dimensional biphoton frequency comb. a,** Illustrative experimental scheme. The signal and idler photons are sent to a 50:50 fiber coupler with orthogonal polarizations for the generation of polarization entanglement. P: polarizer; S.C.: single counts; C.C.: coincidence counts. **b**, Polarization entanglement measurements with P1 fixed at 45° and 90°. In both cases, we measured the coincidence counting rates at the two outputs while changing P2 from 0° to 360°. As shown by the black line (for P1 = 45°) and red line (for P1 = 90°), both measured results fit well with sinusoidal curves, with accidentals-subtracted visibilities of 91.2 % and 93.0 %, respectively.

## VII. High-dimensional hyperentanglement and Bell inequality statistics

We performed a series of hyperentangled measurements of the biphoton frequency comb with the visibilities summarized in Table 1 below, for the different P1 settings and different time-bin settings, across the energy-time basis and the polarization basis. The resulting standard deviation violation of the Bell inequality is computed correspondingly. Measurements are performed at 80-second integration times based on the tradeoff between the setup's long-term stability over the complete hyperentanglement measurements and sufficiently reduced standard deviations of the coincidence counts. The Clauser-Horne-Shimony-Holt (CHSH) $S$ parameter is determined from the polarization analysis angle set of the $(|H\rangle|V\rangle+|V\rangle|H\rangle)$ triplet state:

$$S \equiv |-C(\pi/2, 7\pi/8) + C(\pi/4, 7\pi/8) + C(\pi/4, 5\pi/8) + C(\pi/2, 5\pi/8)| \qquad (31)$$

**Table 1 | Visibilities for the interference fringes in the high-dimensional hyperentanglement measurement and Bell inequality violations.** "st.d." denotes standard deviation $\sigma$.

|  |  | Time bin #0 | Time bin #1 | Time bin #2 | Time bin #3 | Time bin #4 |
|---|---|---|---|---|---|---|
| P1 = 45°, Polarization basis | Visibility V | 82.9% | 80.1% | 77.3% | 73.8% | 77.3% |
|  | V (dark counts subtracted) | 96.3% | 96.2% | 95.9% | 93.8% | 95.8% |
|  | Bell violation (by st.d. $\sigma$) | 8.83 | 9.46 | 10.5 | 6.8 | 9.33 |
| P1 = 45°, Energy-time basis | Visibility V | 81.1% | 74.1% | 68.0% | 60.0% | 44.2% |
|  | V (dark counts subtracted) | 94.1% | 87.5% | 83.3% | 74.7% | 53.2% |
|  | Bell violation (by st.d. $\sigma$) | 7.34 | 5.19 | 4.83 | 2.12 | none |
| P1 = 90°, Polarization basis | Visibility V | 81.4% | 76.6%% | 75.8% | 75.1% | 75.3% |
|  | V (dark counts subtracted) | 96.9% | 95.0% | 95.8% | 94.5% | 95.7% |
|  | Bell violation (by st.d. $\sigma$) | 10.95 | 8.09 | 7.62 | 8.51 | 7.97 |
| P1 = 90°, Energy-time basis | Visibility V | 80.3% | 72.3% | 67.5% | 59.4% | 44.2% |
|  | V (dark counts subtracted) | 95.9% | 89.0% | 84.3% | 73.6% | 54.1% |
|  | Bell violation (by st.d. $\sigma$) | 7.87 | 8.34 | 5.67 | 1.92 | none |
| CHSH $S$ parameter |  | 2.71 | 2.67 | 2.76 | 2.74 | 2.53 |